\documentclass[mathptm
    ,final            
  ]
  {aipproc}

\layoutstyle{8x11single}

\begin{document}

\title{WIMP Gamma Rays From the Galactic Center with GLAST and Accelerator Comparison}


\author{Aldo Morselli}{
  address={INFN and University of Roma Tor Vergata, via della Ricerca Scientifica 1, Roma, Italy}
}

\author{Andrea Lionetto}{
  address={INFN and University of Roma Tor Vergata, via della Ricerca Scientifica 1, Roma, Italy}
}
\author{Eric Nuss}{
  address={LPTA, University of  Montpellier 2, CNRS/IN2P3, Montpellier, France\\ \vspace*{0.5cm}
 Representing the GLAST LAT Collaboration}
}

\begin{abstract}
We will describe the prospects for detecting gamma-rays from WIMP's annihilation in the Galactic Center and we compare this search with the possibilities at LHC and with space antimatter experiments like PAMELA.

\end{abstract}

\maketitle



The space satellite GLAST is expected to play a crucial role in indirect DM searches, thanks both to
its ability to perform observations at energy scales comparable to the mass of common
DM candidates and to its potential of making deep full-sky maps in gamma-rays, thanks to its large ($\sim 2.4$ sr) field-of-view~\cite{glast}.

A theoretically particularly well-motivated type of Weakly Interacting Massive Particle (WIMP) dark matter candidate is the neutralino (see \cite{lsp} for a classic review) that appears in most supersymmetric extensions to the SM as the lightest supersymmetric particle (LSP) and is given by     a linear combination of the superpartners of the gauge and Higgs fields.
 
\subsection{mSUGRA GLAST reach}

We focus 
on the most widely studied neutralino, 
in the most restrictive supersymmetric extension of the Standard Model,
the minimal supergravity (mSUGRA) framework.
We fix the five mSUGRA input parameters:
 $$m_{1/2},\;\;  m_0,\;\;  sign(\mu),\;\;  A_0\;\; \rm{and}\;\; \tan\beta \;,$$
where $m_0$ is the common scalar mass, $m_{1/2}$ is the common gaugino mass and $A_0$ 
is the proportionality factor between the supersymmetry breaking trilinear couplings and the 
Yukawa couplings. $\tan\beta$ denotes the ratio of the vacuum
expectation values of the two neutral components of the SU(2) Higgs doublet, while
the Higgs mixing $\mu$ is determined (up to a sign) by imposing the Electro-Weak Symmetry 
Breaking (EWSB) conditions at the weak scale. The parameters at the weak energy scale 
are determined by the evolution of those at the unification scale, according to the renormalization 
group equations (RGEs). For this purpose, we have made use of the ISASUGRA RGE package in the ISAJET 7.64 software \cite{Baer00}. 
After fixing the five mSUGRA parameters at the unification scale, we extract from the ISASUGRA output
the weak-scale supersymmetric mass spectrum and the relative mixings. Cases in which the
lightest neutralino is not the lightest supersymmetric particle or there is no
radiative EWSB  are disregarded.
The ISASUGRA output is then used as an input in the {\sffamily DarkSUSY} \ package . The latter
is exploited to: 
a) reject models which violate limits recommended
by the Particle Data Group 2002 (PDG)
b) compute the neutralino
relic abundance, with full numerical solution of the density evolution equation
including resonances, threshold effects and all possible coannihilation 
processes \cite{newcoann} 
c) compute the neutralino annihilation rate at zero 
temperature in all kinematically allowed tree-level final states (including fermions,
gauge bosons and Higgs bosons);
d) estimate the induced gamma-ray yield by 
linking to the results of the simulations performed with the Lund Monte Carlo 
program Pythia as implemented in the {\sffamily DarkSUSY} package.

\begin{figure}[ht]
   \includegraphics[height=.27\textheight]{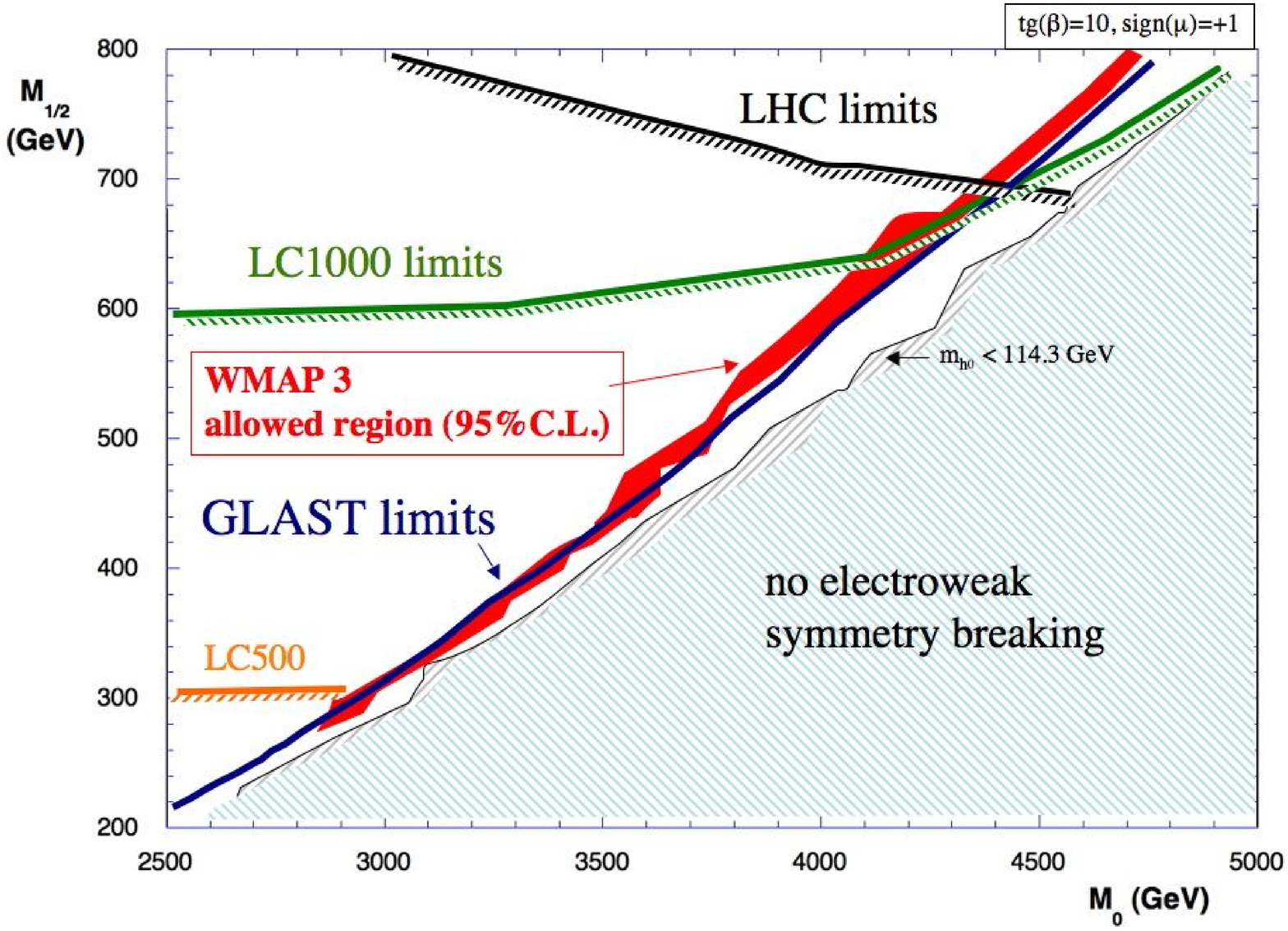}
 \includegraphics[height=.28\textheight]{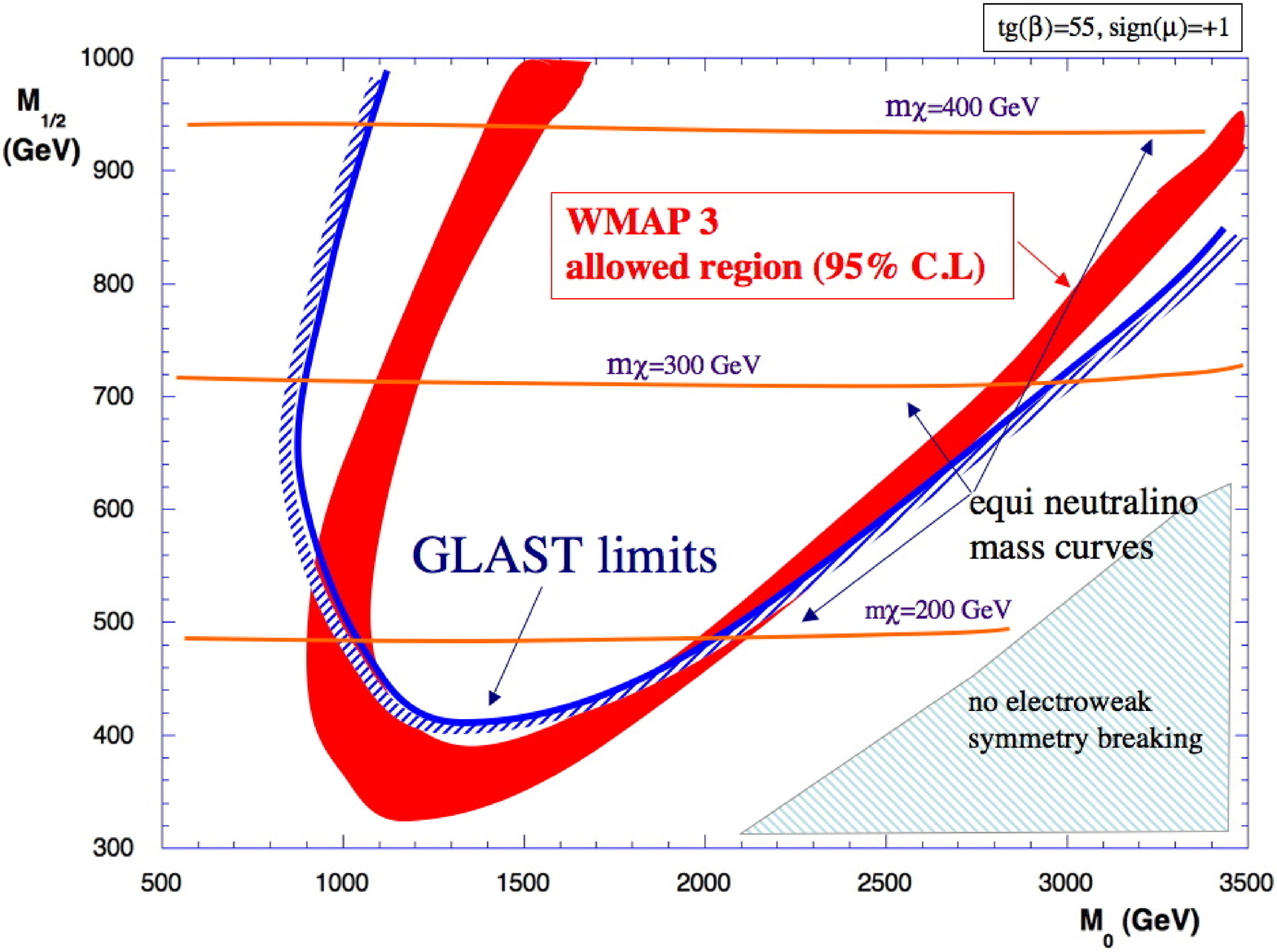}
 \end{figure}
\begin{figure}[ht]

   \includegraphics[height=.26\textheight]{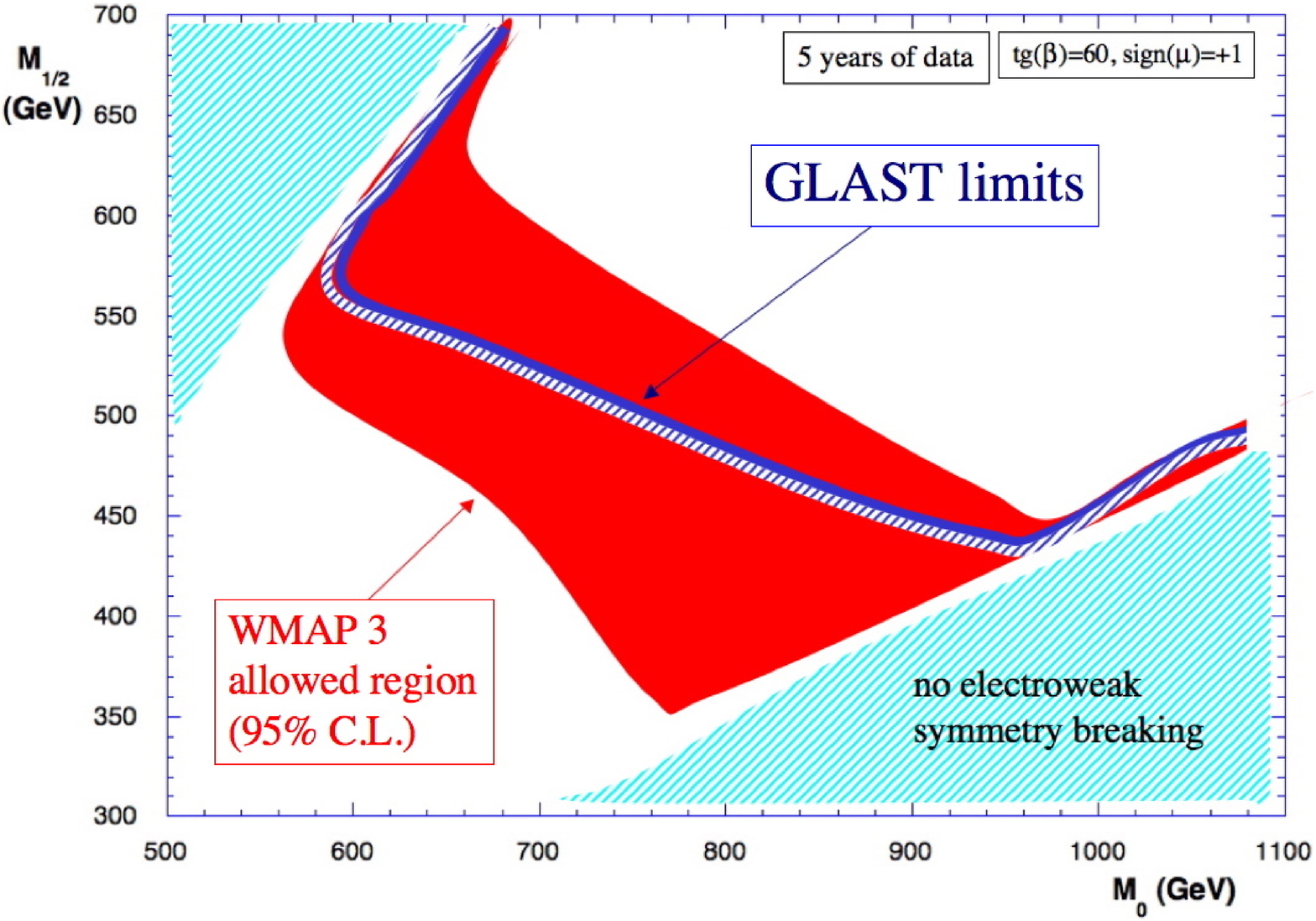}
  \includegraphics[height=.26\textheight]{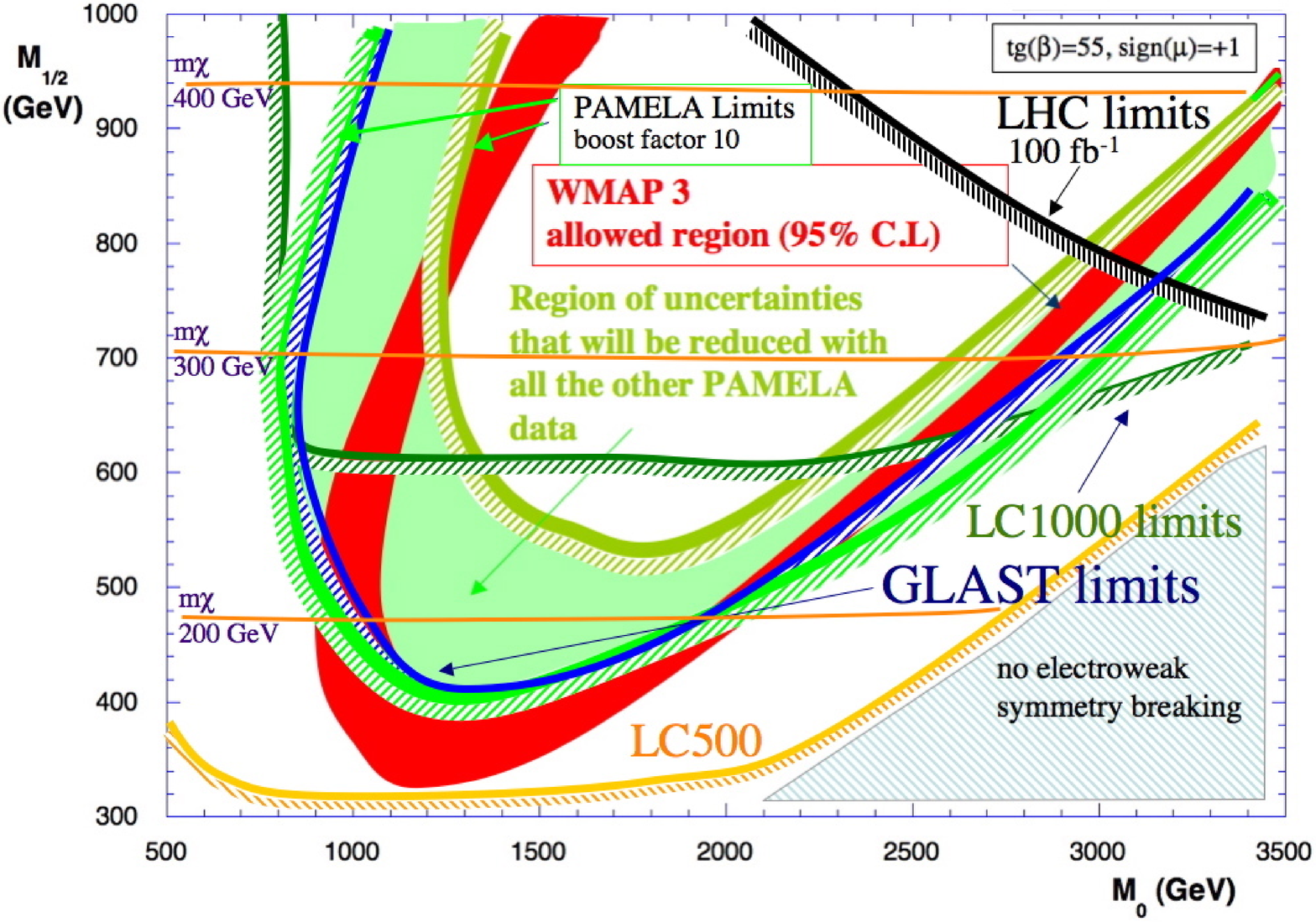}
  \caption{GLAST sensitivity to a dark matter signal via the observation of WIMP annihilation photons (continuum spectrum) in the $m_{1/2}$ and $m_0$ mSUGRA parameter plane for   $\tan\beta=10$,  $55$ and $60$. GLAST 3$\sigma$ sensitivity is shown at the blue line and below.  The lower right plot shows the comparison for $\tan\beta=55$ with LHC, LC  and the antimatter experiment PAMELA. The stripped regions correspond to models that are excluded either by incorrect ElectroWeak Symmetry Breaking (EWSB), LEP bounds violations  or because the neutralino is not the Lightest Supersymmetric Particle (LSP).}
   \label{susyGlast}
\end{figure}

Figure~\ref{susyGlast}  shows our estimates of GLAST sensitivity to a dark matter signal via the observation of WIMP annihilation photons (continuum spectrum) in the $m_{1/2}$ and $m_0$ mSUGRA parameter plane for   $\tan\beta=10$, 55 and 60. These figures have been obtained performing a detailed scan in the mSUGRA parameter space, computing for each model the neutralino induced $\gamma$-ray flux and the relic density. The lower right plot shows the comparison for $\tan\beta=55$ with the exclusion limits from LHC, LC  \cite{Baer} and the antimatter experiment PAMELA \cite{jcap}. The values of the neutralino mass is also shown in both figures on the right.
   For the region in red, the cosmologically allowed WIMP region, the signal  above the blue line ($M_{WIMP} \sim 200 GeV$) is not observable by GLAST due the higher WIMP mass as one moves to higher $m_{1/2}$.
The dark matter halo used for the GLAST indirect search sensitivity estimate is a truncated Navarro Frank and White (NFW) halo profile as used in \cite{dark}.  For steeper halo profiles (like the Moore profile) the GLAST limits move up, covering a wider WMAP \cite{WMAP} allowed region, while for less steep profile (like the isothermal profile) the GLAST limits move down, covering less WMAP allowed region.  

\subsection{Model Independent GLAST Reach}

\begin{figure}
 \includegraphics[height=.28\textheight]{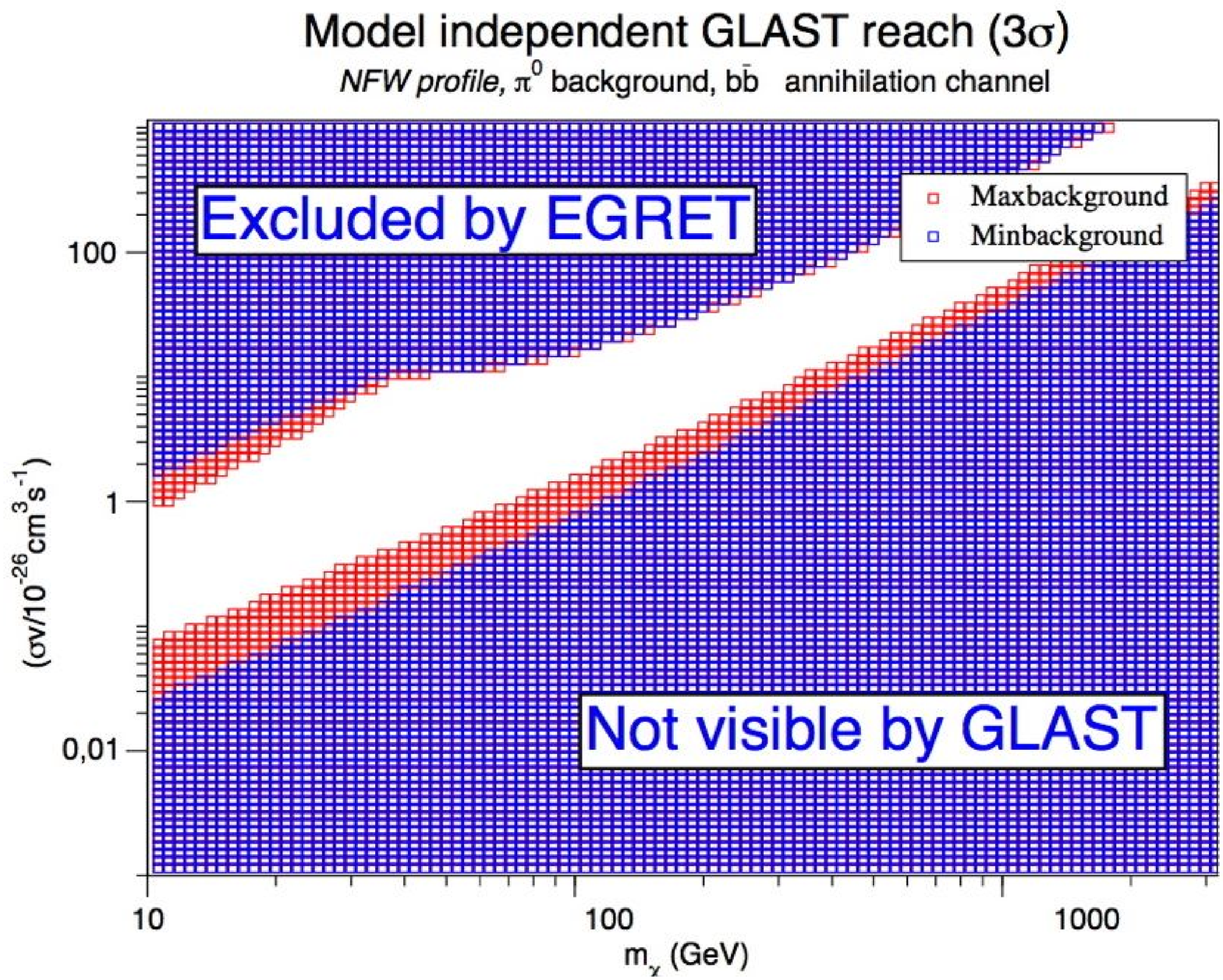}
   \includegraphics[height=.28\textheight]{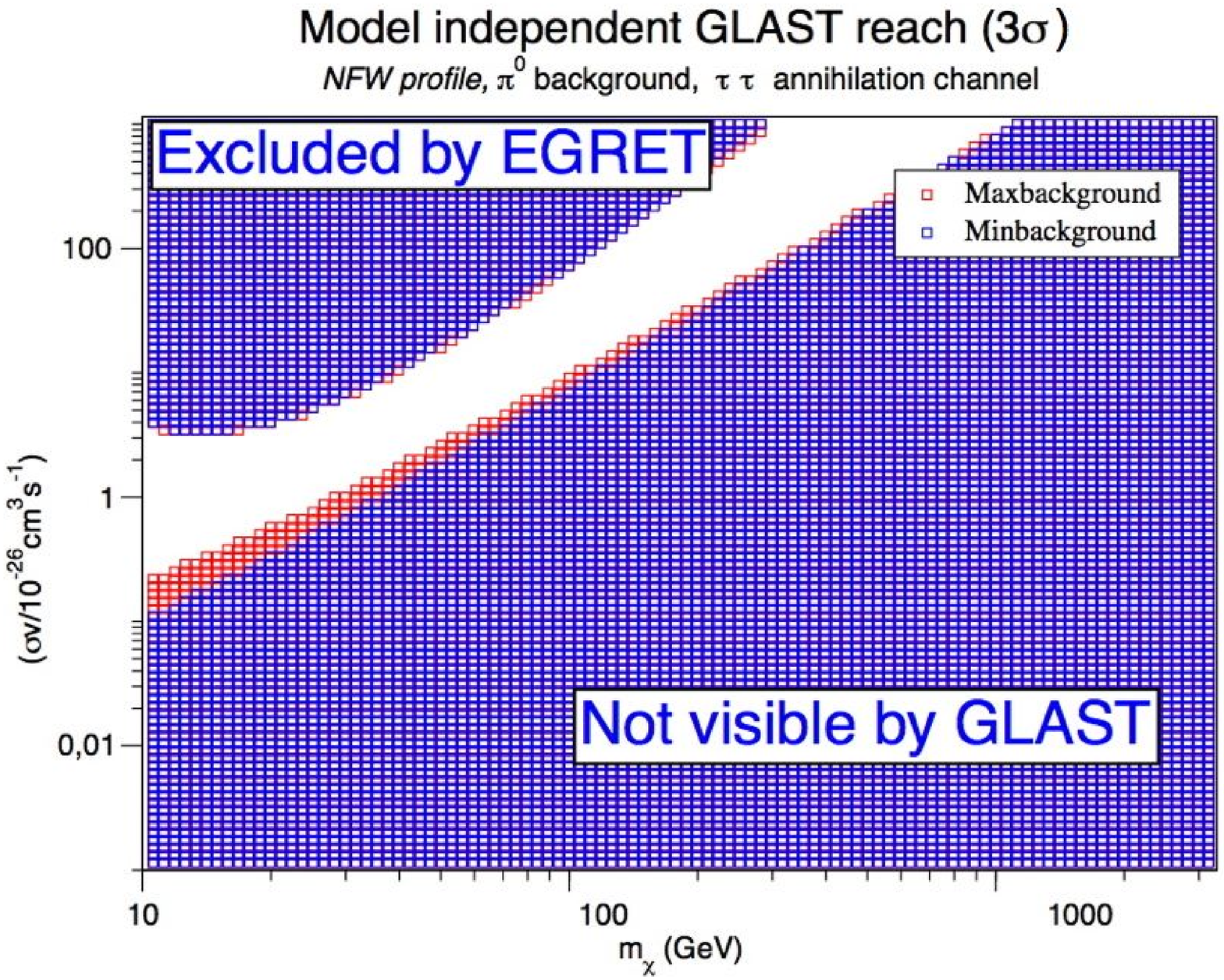}
  \caption{Cross Section times WIMP velocity versus the WIMP mass. The white region is  allowed by EGRET data and detectable by GLAST }     \label{ind}
 \end{figure}

The expression of the $\gamma$-ray continuum flux for a generic WIMP at a given photon energy $E$  is given by
\begin{equation} 
  \phi_{{\rm wimp}}(E)=\frac{\sigma v}{4\pi} \sum_f \frac{dN_f}{dE} B_f
  \int_{l.o.s} dl  \frac{1}{2}\frac{\rho(l)^2}{m_{{\rm wimp}}^2} 
\label{gammafluxcont}
\end{equation}
This flux depends from the WIMP mass $m_{\rm wimp}$, the total annihilation cross
section  times WIMP velocity $\sigma v$ and through the sum of all the photon yield
$dN_f/dE$ per each annihilation channel weighted by
the corresponding branching ratio $B_f$. The flux~(\ref{gammafluxcont}) also depends from the WIMP density in the galactic halo $\rho(l)$. The integral has to be performed along the
line of sight (l.o.s.).
As pointed out in \cite{dark}, apart from the $\tau \bar\tau$  channel,  the photon yields are quite similar. So  
fixing the halo density profile (for example a NFW profile), a dominant annihilation channel (that is $b\bar{b}$, $t\bar{t}$, $W^+ W^-$, ...) and the corresponding yield, it is possible to perform a scan in the plane ($m_{{\rm wimp}}$, $\sigma v$) in order to determine the GLAST reach and the regions that are already excluded by the EGRET data  in in the 2 degrees region  around the
galactic center \cite{dark},  \cite{Mayer}, i.e. the flux predicted by the susy+background model must  not exceed the
total flux predicted from EGRET data.
The result of the scan is given in figure \ref{ind}.
For every couple of values ($m_{{\rm wimp}}$, $\sigma v$) we compute the expected flux~(\ref{gammafluxcont}) and we performed a standard $\chi^2$ statistical analysis to see if GLAST is able to disentangle the WIMP contribution among the standard astrophysical $\pi^0$ background as used in \cite{dark} . The result is given at a $3\sigma$ confidence level. The background uncertanties are reflected in the red regions.
We assumed a total exposure of
$3.7\times 10^{10}\, {\rm cm}^2\, s$,  for a period of 4 years of data taking and an angular resolution (at $10$
GeV) of $\sim 3\times 10^{-5}$ sr as it can be derived from the GLAST LAT performances~\cite{glast-lat-perf}.

\subsection{Conclusions}
We showed the GLAST ability to detect an exotic signal from WIMP's annihilation both in 
 mSUGRA and  in a model indipendent  framework. GLAST will be able to probe a good portion of the parameter space.

\end{document}